\documentstyle[stwol,epsfig]{article}

\def\Journal#1#2#3#4{{#1} {\bf #2}, #3 (#4)}

\def\NPB{{\em Nucl. Phys.} B}
\def\PLB{{\em Phys. Lett.}  B}
\def\PRL{{\em Phys. Rev. Lett.}}
\def\PR{{\em Phys. Rev.}}
\def\PRD{{\em Phys. Rev.} D}
\def\PRC{{\em Phys. Rev.} C}
\def\ZPC{{\em Z. Phys.} C}
\def\etal{{\em et al.}}

\def\mco{\multicolumn}
\def\be{\begin{equation}}
\def\ee{\end{equation}}
\def\bea{\begin{eqnarray}}
\def\eea{\end{eqnarray}}

\newcommand{\id}{\mbox{d}}
\newcommand{\bmat}{\left(\begin{array}}
\newcommand{\emat}{\end{array}\right)}
\newcommand{\mcenter}[2]{\hbox to #1cm{\hss$\displaystyle#2$\hss}}
\newcommand{\G}{g}                         % g
\newcommand{\pola}{\stackrel{\rightarrow}{\Rightarrow}}
\newcommand{\polb}{\stackrel{\rightarrow}{\Leftarrow}}
\newcommand{\polc}{\stackrel{\rightarrow}{\Uparrow}}
\newcommand{\pold}{\stackrel{\rightarrow}{\Downarrow}}
\begin{document}

\title{SPIN STRUCTURE OF THE NUCLEON\footnotemark[1]
}

\author{J.P.NASSALSKI}

\address{So\l tan Institute for Nuclear Studies, ul.Ho\.za 69,
00-861 Warsaw, Poland.}

%%%%%%%%%%%%%%%%%%%%%%%%%%%%%%%%%%%%%%%%%%%%%%%%%%%%%%%%%%%%%%

\twocolumn[\maketitle\abstracts{New experimental results 
on the spin dependent structure functions $g_1$ and $g_2$ which 
are determined from
deep-inelastic scattering experiments at CERN, SLAC and DESY are reported.
These results are used to evaluate the Bjorken sum rule and the singlet
axial charge $a_0$. Results are discussed 
in the framework of next-to-leading
order perturbative QCD. The role of the polarised gluons in 
the interpretation of 
the results on $a_0$ is emphasised. New experiments which aim to determine
gluon polarisation are shortly described.}]

\section{Introduction}
One of the most exciting problems in high energy physics is to describe
the nucleon spin in terms of its partonic constituents: quarks and gluons. 
The experimental tool which is used is deep-inelastic scattering of leptons
on nucleons, from which parton distributions are determined. The theoretical
tool is Quantum Chromodynamics (QCD) which describes the dynamics of 
parton interactions.

The interest in this subject was initialised by the discovery by 
the EMC \cite{emc} that only a small fraction of the nucleon's spin
is carried by quarks. The discovery triggered a large experimental and
theoretical activity which is still pursuing.

The main objectives of the present research are focused on the determination
and interpretations of the singlet axial charge ($a_0$) and the Bjorken sum
rule (BSR), the former related to the nucleon's spin carried by partons and
the latter one regarded as a test of perturbative QCD.

The most recent experimental results are from the SMC at CERN, E142, E143 and E154
at SLAC and HERMES at DESY. There is a large
progress on the theoretical side, where  next-to-leading order (NLO)
perturbative QCD calculations became available. The review of these 
results and of their interpretations is given below.

\section{Determination of $g_1$  and $g_2$}
The spin-dependent structure functions $g_1$  and $g_2$ are determined from
deep-inelastic scattering of polarised electrons and muons on polarised proton,
deuteron and $^3$He targets. The one-photon exchange cross section can be written 
\cite{jaffe} in terms 
of the spin-independent ($\overline{\sigma}$) and the spin-dependent 
cross-section($\Delta \sigma$):
\begin{figure}%%[hpb]
\begin{tabular}{p{0.45\textwidth}}
\vskip-3cm
\epsfig{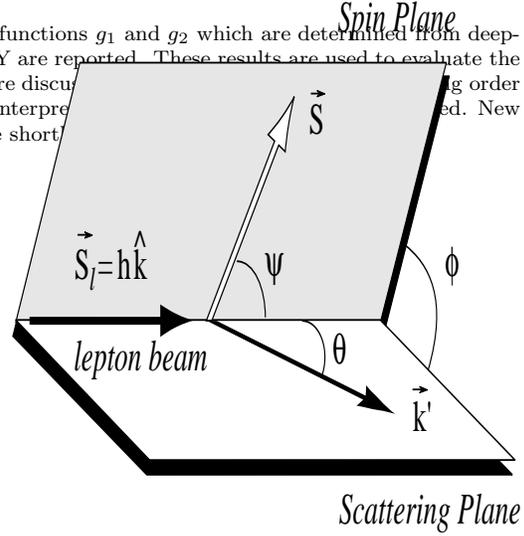}
\vskip-0.3cm
\caption{Definition of angles $\psi$ and $\phi$.}
\end{tabular}
\end{figure}
\begin{equation}
\label{eq:cross_alt}
\frac{\id^3\sigma(\psi)}{\id x\,\id Q^2\,\id\phi} =
     \frac{\id^3\overline{\sigma}}{\id x\,\id Q^2\,\id\phi}
     +
     \frac{\id^3\Delta \sigma (\psi)}{\id x\,\id Q^2\,\id\phi}
\end{equation}

where $x=Q^2/(2m\nu)$ is the Bjorken scaling variable, $m$ is the proton mass, 
$-Q^2$ and $\nu$ are the invariant
mass and the energy of the virtual photon in the laboratory system, respectively.
The angle $\phi$ is  between the lepton scattering plane and the target spin plane
and the angle $\psi$ is between lepton momentum and the target spin as shown in 
Figure~1.
%%\vskip-0.5cm

The two terms can be written as:
\begin{eqnarray}
\lefteqn{\frac{\id^3\overline{\sigma}}{\id x\,\id Q^2\,\id\phi} = 
              \frac{2\alpha^2}{Q^4}\cdot} \nonumber \\
              & \left\{y^2 \left(1-\frac{2m_l^2}{Q^2}\right) F_1+
              \frac{1}{x}\left(1-y-\frac{y^2\gamma^2}{4}\right) F_2 \right\} &,  \\  
\lefteqn{\frac{\id^3\Delta \sigma(\psi)}{\id x\,\id Q^2\,\id\phi} =
              \frac{4\alpha^2}{Q^4} y \cdot} \nonumber \\
              & \left\{
                   \cos\psi\left[\left(1-\frac{y}{2}-\frac{y^2\gamma^2}{4}\right)
                   g_1-\frac{y}{2}\gamma^2g_2\right]\right. & \nonumber \\
              &  \left.-\gamma\sin\psi
                   \sqrt{ 1-y-\frac{y^2\gamma^2}{4} }\,
                   \left(\frac{y}{2}g_1+g_2\right)\cos\phi\right\}. &
\end{eqnarray}
Here $F_1$ and $F_2$ are spin-independent structure 
functions, $y=\nu/E$, where E
is incident lepton energy, $m_l$ is the lepton mass and $\gamma^2=Q^2/\nu^2=4x^2m^2/Q^2$
vanishes at large $Q^2$.
The experiments use longitudinally polarised (polarisation $P_b$) leptons and 
measure asymmetries of 
the cross-sections for opposite orientations of target polarisations (polarisation $P_t$):
\begin{equation}
\label{eq:asym}
A_{\|(\bot)}=\frac{\sigma^{\polb(\polc)}-\sigma^{\pola(\pold)}}
                    {\sigma^{\polb(\polc)}+\sigma^{\pola(\pold)}} fP_bP_t
            =\frac{\Delta\sigma_{\|(\bot)}}{2\cdot\bar\sigma} fP_bP_t,
\end{equation}
where $\rightarrow(\Rightarrow)$ denote the beam (target) polarisations and $\bar\sigma$
refers to spin-averaged cross-section. The symbols $\|$ and $\bot$ indicate longitudinal 
($\psi=0$ or $\pi$) and transverse ($\psi=\pi/2$) orientations of the target spin, 
respectively. In the
transverse case the asymmetry  is measured between interactions where the scattering 
plane is at an angle $\phi$ and $\pi\pm\phi$ (see Fig.1). The dilution factor $f$ 
accounts for the interactions on the unpolarisable material
within the target fiducial volume.

These asymmetries can be written in terms of $g_1$ and $g_2$ as \\
\begin{array}[b]{lrlll}
\normalsize
A_{\|}   = [ & \frac{g_1-\gamma^2 g_2}{F_1} & +\eta\gamma  
                         & \frac{g_1+g_2}{F_1}\gamma    & ]DfP_bP_t \\
A_{\bot} = [-& 
               \underbrace{\frac{g_1-\gamma^2 g_2}{F_1}}_{A_1} & \xi\gamma+ 
                         & 
               \underbrace{\frac{g_1+g_2}{F_1}\gamma}_{A_2} & ]dfP_bP_t 
\end{array}
where $\eta$, $\xi$, $D$ and $d$  are kinematic factors and $A_1$ and $A_2$
are virtual photon asymmetries \cite{jaffe}. Therefore
\[ \left.\begin{array}{ccc}
g_1    & = & (A_1+\gamma A_2)/(1+\gamma^2) \\
g_1+g_2& = & A_2/\gamma 
   \end{array} \right\} \cdot F_1 \]
where 
\[ F_1(x,Q^2)=\frac{(1+\gamma^2)F_2(x,Q^2)}{2x(1+R(x,Q^2)}. \]
The experiments
use the NMC parameterisation  \cite{nmcf2} for $F_2$ and SLAC parametrisation
\cite{slacr} for $R=\sigma_L/\sigma_T$, the ratio of photoabsorption
cross-sections.

\section{Recent experiments.}
Recent experiments on polarised deep-inelastic scattering are listed in Table 1.
Table 2 indicates distributions which have been determined and also the derived quantities evaluated at 
fixed values of $Q_0^2$; the first moments $\Gamma_1=\int_0^1{dx g_1(x,Q_0^2)}$,
twist-3 matrix elements $d_k$ and the polarised valence ($\Delta u_v,\Delta d_v$) and
sea ($\Delta \overline q$) quark distributions. The semi-inclusive asymmetries will
be discussed in Sec.10. 
%\vskip-4cm
\begin{table*}[t]
\begin{center}
\caption{Recent experiments on polarised DIS.}
\vspace{0.4cm}
\begin{tabular}{|l|cc|cr|l|c|c|} \hline
Expt. & Beam & (GeV) & \multicolumn{2}{c|}{Target} & P$_b$ & P$_t$ & f$^{(a)}$ \\ 
 & & & & & & & \\ \hline
SMC   & $\mu$& 190       & C$_4$H$_9$OH          & p & 0.8 & 0.9     & 0.12 \\ 
      &      &190$^{(b)}$& C$_4$D$_9$OH          & d & 0.8 & 0.5     & 0.20 \\
\em in 1996:&& 190       & $^{14}$NH$_3$         & p & 0.8 & 0.9     & 0.17 \\ \hline
E142  & $e$  & 19-25     & $^3$He                & n & 0.4 & 0.3-0.4 & 0.11 \\
E143  &      & 10-29     & $^{15}$NH$_3$         & p & 0.8 & 0.7     & 0.13-0.17 \\
      &      & 10-29     & $^{15}$ND$_3$         & d & 0.8 & 0.3     & 0.22-0.25 \\
E154  &      & 49        & $^3$He                & n & 0.8 & 0.4     & 0.11 \\
E155  &      & 49        & $^{15}$NH$_3$         & p & 0.8 & 0.7     & 0.13-0.17 \\
\em in 1997:&& 49        & $^{15}$ND$_3$ (or LiD)& d & 0.8 & (0.5)   & (0.5) \\ \hline
HERMES& $e$  & 28        & $^3$He                & n & 0.5 & 0.5     & 1/3 \\
\em in 1996:&& 28        & H$_2$                 & p & 0.5 & 0.9     & 1 \\ \hline
\mco{8}{l}{$^{(a)}$ effective $f$ in the target fiducial volume.} \\
\mco{7}{l}{$^{(b)}$ includes also data at 100 GeV.} \\
\end{tabular}
\end{center}
%\end{table*}
%\vskip-8cm
%\begin{table*}%%[t]
\begin{center}
\caption{Distributions determined from polarised DIS.} 
\vspace{0.4cm}
\begin{tabular}{|c|c|l|} \hline
Distribution & Derrived quantity & Experiment \\
 & & \\ \hline
$g_1^{p(d)}(x,Q^2)$     & $\Gamma_1^{p(d)}$   & SMC, E143 \\ \hline
$g_1^{n}(x,Q^2)$        & $\Gamma_1^{n}$      & \em E142 - being reevaluated \\  
                        &                     & \em E154 (meas. range) \\
                        &                     & HERMES \\
                        &                     & \em SMC and E143 - from (p,d) \\ \hline
$g_2^{p(d)}(x)$         & $d_{k=2,4,6}^{p(d)}$& E143 \\
                        &                     & \em SMC - $A^{p(d)}_{\bot}$ only \\ \hline
\em semi-inclusive       & $\Delta u_v(x)$      & SMC \\
\em asymmetries:        & $\Delta d_v(x)$      & \\
$\left(A_{\|}\right)^{\pi^+(\pi^-,\pi^+-\pi^-)}(x)$                       
                        & $\Delta \overline q(x)$ & \\ \hline
\end{tabular}
\end{center}
\end{table*}
%%\newpage

\section{Interpretation of $g_1$ in the QCD-improved parton model}
In the QCD-improved parton model, $g_1$ is expressed in terms of distribution
functions 
of longitudinal polarisation of partons; quarks $(\Delta q_i)$ and antiquarks 
$(\Delta \overline{q_i})$ of flavour $i$ and gluons $(\Delta g)$. Here 
$\Delta f \equiv f^+ - f^-$ and $f^{+(-)}$ are 
distribution functions of partons with
spin parallel (antiparallel) to nucleon spin: 
\begin{eqnarray}
g_1(x,Q^2) = && \hskip-5mm \frac{1}{2}\sum_{i=1}^{n_f} e_i^2C_i(x,Q^2)\otimes
\left[\Delta q_i + \Delta \overline {q_i} \right](x,Q^2) \nonumber \\
&& \hskip-5mm + C_g(x,Q^2) \otimes \Delta g(x,Q^2),
\end{eqnarray}
where  $n_f$ is the number of active flavours and $C_{i(g)}$ are quark (gluon) 
coefficient functions and
$C \otimes \Delta f \equiv \int_{x}^{1} \frac{{\rm d}y}{y} C(\frac{x}{y},Q^2)
\Delta f (y,Q^2)$.

The coefficient functions can be expanded in powers of $\alpha_s$ and at
next-to-leading order (NLO) they are:
\begin{eqnarray}
C_i(x,Q^2)  = & \delta (x-1) & + \frac{\alpha_s}{2\pi} C_i^{(1)}(x) + {\cal O}(\alpha_s^2) \nonumber \\
C_g(x,Q^2) =  &   0          & + \frac{\alpha_s}{2\pi} C_g^{(1)}(x) + {\cal O}(\alpha_s^2) \nonumber
\end{eqnarray}
\begin{equation}\end{equation}
At leading order (LO) we therefore get \\
\mbox{$g_1(x) = \frac{1}{2} \sum_{i}^{} e_i^2
(\Delta q_i(x) + \Delta \overline {q_i}(x))$}, which is simply 
related to $\Delta q = q^+ - q^-$.
This can be compared to the spin-averaged structure function 
$F_1(x) = \frac{1}{2} \sum_{i}^{} e_i^2 (q_i(x) + \overline {q_i}(x))$ 
related to
the sum $q=q^+ + q^-$.

\section{New results on $g_1$ and $g_2$}
\subsection{Results on $g_1^p$}

Figure~2 shows results on $g_1^p(x)$ from the experiment
E143 \cite{e143p} and from an updated (preliminary) analysis of the SMC data
\cite{SMCp}. The points are shown at measured $Q^2 > 1\,{\rm GeV}^2$. In 
the overlap region the results agree within the errors. At smaller values
of $x$ the SMC points have large errors due to the cut $Q^2 > 1\,{\rm GeV}^2$
which eliminates most of the events.
Figure~3 shows the corresponding 
asymmetry $A_1^p = g_1^p/F_1^p$ and also the one (preliminary) 
obtained with the cut $Q^2 > 0.2\,{\rm GeV}^2 $; in the latter case the $x$ range 
is extended down to $x = 0.001$ and the results do not show any large 
variation with $x$.

Figure~4 shows the $Q^2$-dependence of the asymmetry $A_1^p(x,Q^2)$ from 
the E143, the SMC and the EMC \cite{emc} in bins of $x$. Within the errors 
no $Q^2$-dependence is seen between the data sets from SLAC and CERN, 
covering different ranges of $Q^2$ at fixed $x$.

\subsection{Results on $g_1^d$}
Figure~5 shows results on $g_1^d(x)$ from the experiment
E143 \cite{e143d} and from  the SMC (preliminary)
\cite{SMCd}. The points are shown at measured $Q^2 > 1\,{\rm GeV}^2$. In 
the overlap region the results agree within the errors.

The $Q^2$-dependence of the asymmetry \\
$A_1^d(x,Q^2)$  obtained from 
the E143 and the SMC 
data in the region $Q^2 > 0.2\,{\rm GeV}^2$ 
is shown in Figure~6.
Within the errors no variation of the asymmetry $A_1^d$ 
with $Q^2$ is observed. Note however, that at the smallest values of $x$
and $Q^2$ covered by the E143, their analysis \cite{E143_qdep} based on
all previously published data on $A_1^p(x,Q^2)$ and $A_1^d(x,Q^2)$ 
indicates a behaviour $A_1^{p(d)}(x,Q^2) \sim (1 + C^{p(d)}(x))/Q^2$.  
%\newpage
%\vskip-3cm
\begin{figure}%[t]
%\begin{tabular}{p{0.45\textwidth}}
\vskip-4.2cm
\epsfig{file=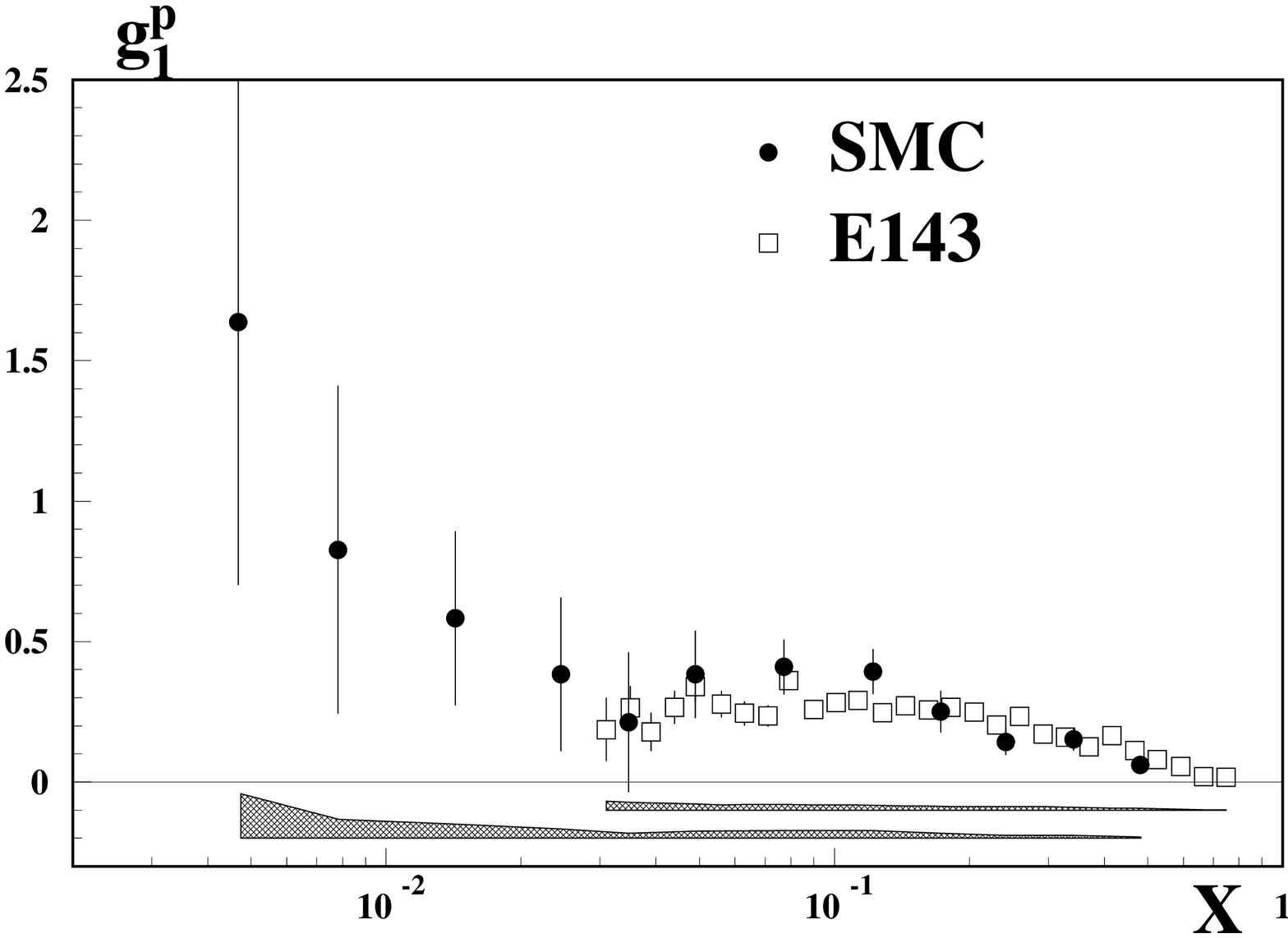,height=6.0cm,width=\linewidth,clip= ,bbllx=0pt,bblly=0pt,bburx=650pt,bbury=450pt} 
%\epsfig{file=fig_3.eps,width=\linewidth}
%\vskip-0.3cm
\caption{The structure function $g_1^p(x)$ from the SMC (preliminary)
and from the E143. Points are shown at measured 
$Q^2 > 1\,{\rm GeV}^2$ with the statistical errors. The sizes of the systematic errors 
are indicated with the bands.}
%\end{tabular}
%\end{figure}
%\vskip-0.5cm
%\begin{figure}%%[h]
%\begin{tabular}{p{0.45\textwidth}}
%\vskip-1.5cm
%\epsfig{file=fig_4.ai,height=6.0cm,width=\linewidth,clip=
%,bbllx=0pt,bblly=0pt,bburx=650pt,bbury=450pt} 
%\epsfig{file=fig_4.ai,height=6.0cm,width=\linewidth}
\epsfig{file=fig_4.ai,width=\linewidth}
%\vskip-0.3cm
\caption{The SMC results on the asymmetry $A_1^p = g_1^p/F_1^p$ 
for $Q^2 > 1\,{\rm GeV}^2$ and $Q^2 > 0.2\,{\rm GeV}^2 $ (preliminary).The errors are statistical. The size of the systematic errors for the result at
$Q^2 > 0.2\,{\rm GeV}^2 $ is indicated with the band.}
%\end{tabular}
\end{figure}
%\vskip-0.5cm

\subsection{Results on $g_1^n$}
The structure function $g_1^n$ is obtained from the measurements of $g_1^d$
and $g_1^p$ (SMC, E143) or from $g_1^{^3He}$ (E142, E154, HERMES).
In both cases different corrections are needed, however all data on $g_1^n$
are consistent, as will be shown below.

\begin{figure}%%[h]
\begin{tabular}{p{0.45\textwidth}}
%\vskip-0.5cm
\epsfig{file=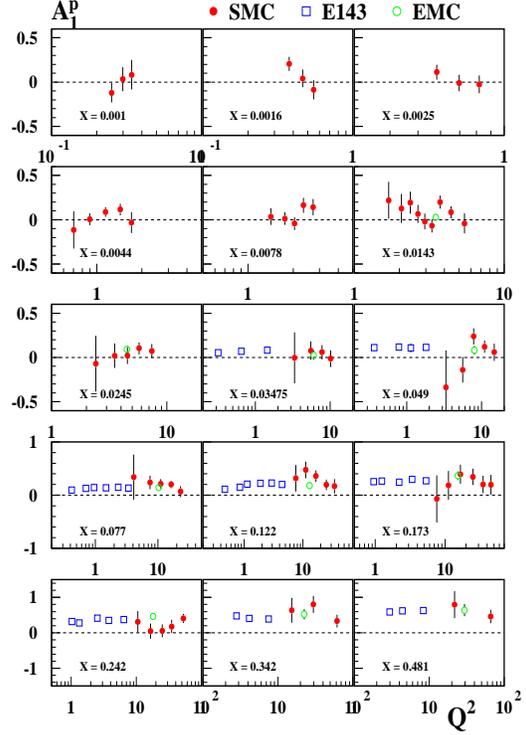,height=10.0cm,width=\linewidth,clip=,bbllx=20pt,bblly=80pt,bburx=550pt,bbury=750pt}
%\epsfig{file=fig_5a.ps,height=6.0cm,width=\linewidth}
%\vskip-0.3cm
\caption{$Q^2$ dependence of the asymmetry $A_1^p(x,Q^2)$ from 
the E143, the SMC and the EMC in bins of $x$. The SMC results
are preliminary. Only statistical
errors are shown. }
\end{tabular}
\end{figure}
%\vskip-0.5cm

\begin{figure}%%[h]
\begin{tabular}{p{0.45\textwidth}}
\vskip-1.2cm
\epsfig{file=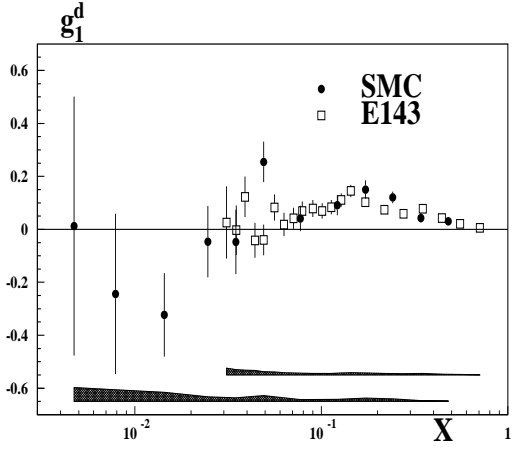,height=6.0cm,width=\linewidth,clip=
,bbllx=0pt,bblly=0pt,bburx=650pt,bbury=450pt} 
%\vskip-0.3cm
\caption{The structure function $g_1^d(x)$ from the SMC at CERN 
and from the E143 experiment at SLAC. The SMC results are preliminary. 
Points are shown at measured 
$Q^2 > 1\,{\rm GeV}^2$ with the statistical errors. The sizes of the systematic errors 
are indicated with the bands.}
\end{tabular}
\end{figure}
%\vskip-0.5cm

\begin{figure}%%[h]
\begin{tabular}{p{0.45\textwidth}}
\vskip-0.2cm
\epsfig{file=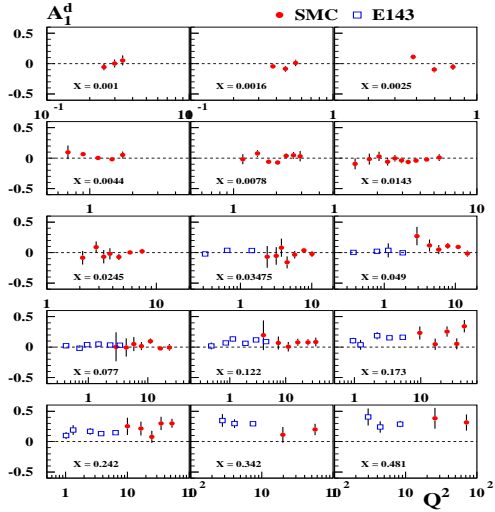,height=6.0cm,width=\linewidth}
%\vskip-0.3cm
\caption{$Q^2$ dependence of the asymmetry $A_1^d(x,Q^2)$ from 
the SMC and the E143 in bins of $x$. SMC Results are preliminary. Only statistical
errors are shown. }
\end{tabular}
\end{figure}
%\vskip-0.5cm

{\underline {Results from deuteron and proton data .}}

In the deuteron, neutron and proton are either in S-state, with both spins
in the direction of the deuteron spin, or in D-state where they are opposite
to it. The probablility of the D-state is 5 $\pm$ 1\% \cite{Dwave} and for the
asymmetries  the correction is 8\%: $A_1^{p+n}=1.08 A_1^d$. The off-mass shell
effects are negligible for $x < 0.7$ \cite{kulagin}. Shadowing effects at small
$x$ were found to give \cite{khan}  $A_1^{p+n}=1.02 A_1^d$ at $x=0.003$; there are no 
independent estimates and this correction is not made. Therefore we have 
\mbox{1.08 $g_1^d = (g_1^p+g_1^n)/$2}.

Figure~7 shows the results for $g_1^n(x)$ from the experiments
E143 and SMC (preliminary). In the region of overlap they are consistent, 
as expected 
from the agreement on $g_1^p$ and $g_1^d$. In the region of small $x$,
covered by the SMC data, the spin-dependent structure functions $g_1^n(x)$
and $g_1^p(x)$ have different signs. In the same region the spin-averaged
structure function $F_2^n(x)$ and $F_2^p(x)$ are known \cite{ratio} to agree 
within $\sim 3\%$ which is interpreted as due to predominance
of the sea quarks. The difference observed here might indicate a relatively important
contribution from polarised valence quarks at small $x$ (see sec.10).

\begin{figure}%%[h]
\begin{tabular}{p{0.45\textwidth}}
%\vskip-1.5cm
%\epsfig{file=fig_7.eps,height=10.0cm,width=\linewidth,clip=
%,bbllx=20pt,bblly=80pt,bburx=550pt,bbury=750pt} 
%\epsfig{file=fig_9.eps,height=6.0cm,width=\linewidth}
\epsfig{file=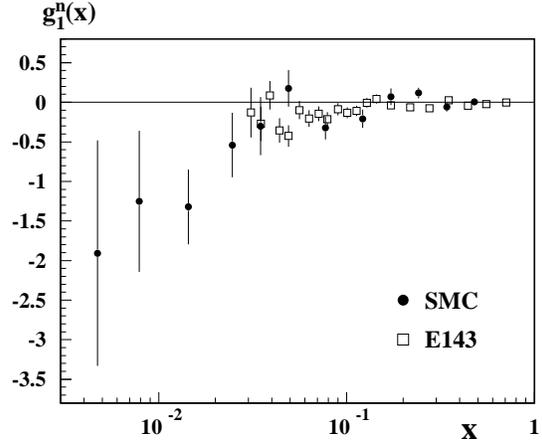,height=6.3cm}
%\vskip-0.3cm
\caption{The structure function $g_1^n(x)$ obtained from proton
and deuteron data of E143 and SMC (preliminary). The error bars show
statistical errors only.}
\end{tabular}
\end{figure}
%\vskip-0.5cm

{\underline {Results from helium data.}}

In the helium wave function the main configurations are: S-wave with two proton
spins in the opposite directions and the neutron spin in the direction of helium 
spin (87\%) and  D-wave with neutron spin opposite to it and the two proton spins parallel (3\%). No other nuclear corrections are made and  
$g_1^{^3He} = ($0.87$g_1^n - $0.03$g_1^p)$/3. However, it has been indicated
\cite{frankfurt} that the effects from an admixture of $\Delta$-isobars 
and from shadowing might be important both for $^3$He and for $^3$H. Nuclear
effects in unpolarised $^3$He will be estimated from measurements of 
$\sigma^{^3He}/(\sigma^d+\sigma^p)$ by HERMES this year \cite{wander}.

Figure~8 shows new, preliminary results from the experiment
E154 at SLAC \cite{e154}. In the same figure preliminary results from the reanalysed
experiment E142 are also included. There is a good agreement between
the two experiments. Results from E154 have very high statistical precision,
comparable or better than the systematic one.

\begin{figure}%%[h]
\begin{tabular}{p{0.45\textwidth}}
\vskip-3.0cm
\epsfxsize=7.7cm
\epsfysize=10cm
\hfil
\epsffile[40 170 530 640]{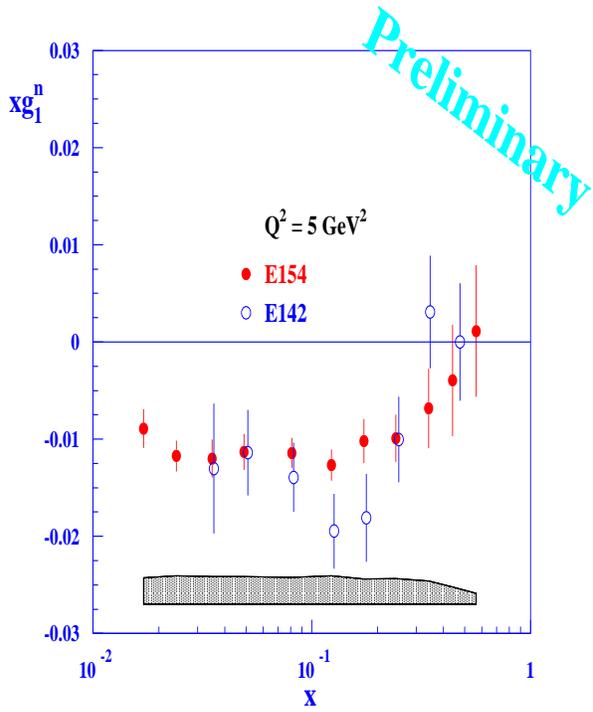}
\hfil
%\vskip-0.3cm
\caption{Preliminary results on the structure function $xg_1^n(x)$ from
the SLAC experiments E154 and E142 (reanalysed), evaluated at 
$Q^2=10\,{\rm GeV}^2$. The error bars show
the sizes of statistical errors. The systematic errors of E154 results
are shown with the band.}
\end{tabular}
\end{figure}
%\vskip-0.5cm

New, preliminary results from HERMES experiment \cite{wander} are shown in
Figure~9.

\begin{figure}%%[h]
\begin{tabular}{p{0.45\textwidth}}
\vskip-2.0cm
\epsfig{file=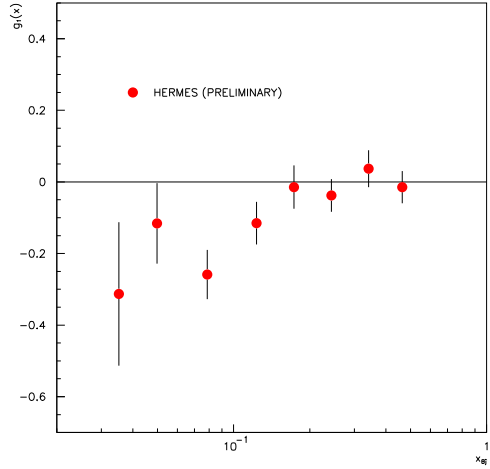,width=\linewidth}%%,width=\linewidth
%%,clip=
%%,bbllx=5pt,bblly=5pt,bburx=530pt,bbury=530pt} 
%\vskip-0.3cm
\caption{Preliminary results on the structure function $xg_1^n(x)$ from
HERMES experiment at DESY. Only systematical errors are shown with
the error bars.}
\end{tabular}
\end{figure}
%\vskip-0.5cm

\section{Results on $g_2^{p(d)}(x)$}

Results on $g_2^p(x)$ and $g_2^d(x)$ come from experiment E143 \cite{g2pd}
and they are
shown in Figure~10. The results are given at measured 
$Q^2 > 1 \,{\rm GeV}^2$. Assuming the validity of Burkhard-Cottingham 
sum rule \cite{bur_cot}, 
\mbox{$\int_{0}^{1} g_2(x,Q^2) {\rm d}x = 0$}, $g_2$ can be written
as a sum of the twist-2 term ($g_2^{WW}$) and a pure twist-3 one 
($\overline{g_2}$):~
\mbox{$g_2(x,Q^2) = g_2^{WW}(x,Q^2) + \overline{g_2}(x,Q^2)$}.
The term $g_2^{WW}$ can be expressed by $g_1$ (Wandzura and
Wilczek \cite{ww}): 

\begin{equation}
g_2^{WW}(x,Q^2) = -g_1(x,Q^2) + \int_{x}^{1} \frac {g_1(z,Q^2)}{y}{\rm d}y.
\end{equation}

%\begin{figure}[h]
%\begin{tabular}{p{0.45\textwidth}}
%\vskip-0.5cm
%\epsfig{file=fig_12.eps,height=10.0cm,width=\linewidth
%,clip=
%,bbllx=110pt,bblly=220pt,bburx=500pt,bbury=650pt} 
%\vskip-0.3cm
%\caption{The structure functions $g_2^p(x)$ and $g_2^d(x)$ from the experiment
%E143.The error bars show the statistical errors. The size of the systematic
%errors is shown with the band. See text for the explanation of the lines.}
%\end{tabular}
%\end{figure}
%\vskip-0.5cm

\begin{figure}
   \begin{center}
      \mbox{\epsfxsize=0.8\hsize\epsfbox{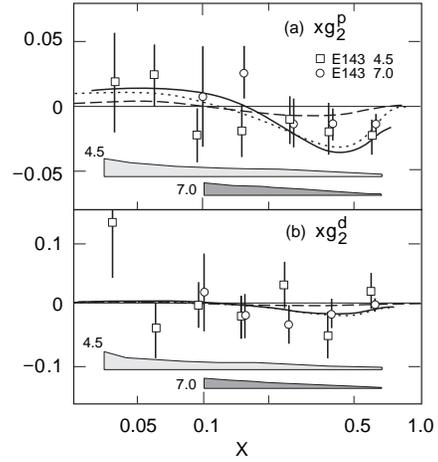}}
      \end{center}
\caption{The structure functions $g_2^p(x)$ and $g_2^d(x)$ from the experiment
E143.The error bars show the statistical errors. The size of the systematic
errors is shown with the band. See text for the explanation of the lines.}   
\end{figure}

The solid line in the figure represents $g_2^{WW}$. It is consistent with
the data indicating that twist-3 term is small.

At smaller values of $x$ there are measurements of $A_2^{p(d)}(x)$ from the SMC
\cite{a2smc}. Results on $A_2^{p}(x)$ are shown in Figure~11 
together with the results from the E143 experiment.
\begin{figure}%%[h]
\begin{tabular}{p{0.45\textwidth}}
\vskip-3.2cm
\epsfig{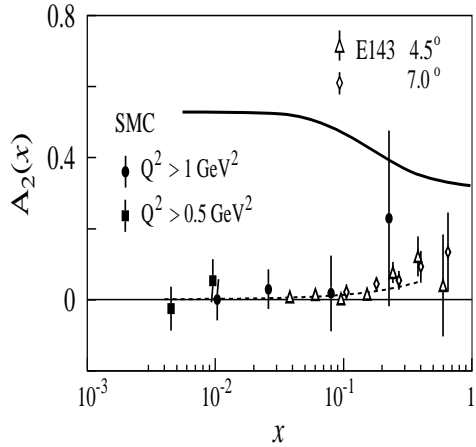}
%,clip=
%,bbllx=5pt,bblly=5pt,bburx=530pt,bbury=530pt} 
%\vskip-0.3cm
\caption{The asymmetry $A_2^p(x)$ from the SMC and from the E143. Only
statistical errors are shown. See text for the explanation of the line.}
\end{tabular}
\end{figure}
%\vskip-0.5cm
Also shown are the SMC results obtained with the cut $Q^2 > 0.5 \,{\rm GeV}^2$.
The solid line in the figure represents an upper limit of $A_2(x)$ given
by $\sqrt{R}$. The present measurements fall substantially below this limit.

\section{The first moment $\Gamma_1(Q_0^2)$}
The main physics interest is in the first moments of the the spin-dependent
structure functions $g_1$:
\begin{equation}
\Gamma_1(Q_0^2)=\int_{0}^{1}g_1(x,Q_0^2){\rm d}x
\end{equation}
where $\Gamma_1$ is the leading-twist (LT) term in the operator
product expansion.

In the experiments $g_1$ is determined in a finite range of $x$, where
$x$ and $<Q^2(x)>$ are correlated. Therefore calculation
of $\Gamma_1$ involves extrapolation of measured $g_1(x,Q^2)$ to 
a common $Q_0^2$ and also extrapolation to $x=0$ and $x=1$.
Finaly, in order to evaluate the LT contribution to $\Gamma_1$,
corrections due to higher-twist terms are required. 

\subsection{Theoretical interpretation}
The first moment of $g_1(x,Q^2)$ can be written as
\begin{eqnarray}
\Gamma_1(Q^2)&=&\frac{1}{2}\left[C_{NS}(Q^2) a_{NS}+ \right. \nonumber \\
              && \left. <e^2>C_S(Q^2)a_0(Q^2)\right]
\end{eqnarray}
in terms of the singlet (S) and non-singlet (NS) axial charges $a_i$,
\[ms^{\mu}a_i=<target,p,spin \mid J_{5i}^{\mu} \mid target,p,spin>,\]
of quark axial-vector current, \mbox{$J_{5i}^{\mu}=\overline{\psi _i}\gamma^{\mu}
\gamma_5\psi _i$}, and Wilson coefficient functions $C_{S(NS)}$
where $p$ is the target momentum. In 
the Quark-Parton Model (QPM) the axial charge of a quark of flavour
$f$ is expressed in terms of the integral of the corresponding distribution
functions of quark polarisations:
$ \Delta q_f(Q^2)\equiv\int_{}^{}\Delta q_f
(x,Q^2){\rm d}x\equiv a_f$.\\
The non-singlet axial charge, 
\[a_{NS}=\sum_{i=1}^{n_f}(e_i^2-<e^2>)a_i),\] 
is conserved and, assuming SU(3)$_f$ symmetry,
it can be written  in terms of $F$ and $D$ constants in hyperon $\beta$-decays:
\begin{eqnarray}
a_{NS}&=&\frac{1}{6} \underbrace{(a_u-a_d)}_{a_3= \mid g_A \mid = F+D}+
       \frac{1}{18}\underbrace{(a_u+a_d-2a_s)}_{a_8=3F-D}+ \nonumber \\
      && \underbrace{(a_u+a_d+a_s-3a_c)}_{if~Q^2>m_c^2}.
\end{eqnarray} 
The singlet axial charge $a_0(Q^2)$ is not conserved because at two loops 
the gluon operator mixes with divergence of the singlet axial-vector current
(this is called "axial anomaly"). As a result, the decomposition
of $a_0$ depends on the renormalisation scheme: 

\[ a_0(Q^2) = \left\{ \begin{array}{ll}
        in~\overline {\rm{MS}}~scheme: \\
        \sum_{i}^{}(\Delta q_i + \Delta \overline{q_i})(Q^2) =\\
          \Delta u_v + \Delta d_v + 2\sum _{i}^{}\Delta \overline{q_i}(Q^2) \\
          \\
        in~Adler-Bardeen~(\rm{AB})~scheme:\\
        \sum_{i}^{}(\Delta q_i + \Delta \overline{q_i}) -
          n_f \frac{\alpha _s(Q^2)}{2\pi} \Delta G(Q^2),
                        \end{array}
                \right. \]
\begin{equation} \end{equation}
where $\Delta G(Q^2)=\int_{}^{}\Delta g(x,Q^2){\rm d}x$ is the integrated gluon
polarisation.
The gluon polarisation enters explicitly in AB scheme \cite{bfr} while in 
$\overline {\rm{MS}}$ scheme  it contributes via polarised sea 
$\Delta \overline{q_i}(Q^2)$.

\subsection{Evolution of $g_1$ with $Q^2$}
The structure function $g_1$ (Eq.5) can be expressed in terms of
the singlet and non-singlet polarised parton distribution functions:
\begin{eqnarray}
g_1(x,Q^2)&=&\frac{<e^2>}{2}\left[C_{NS}\otimes \Delta q_{NS}+\right.
                                                           \nonumber \\
          & &\left. C_S\otimes \Delta q_S+2n_fC_g\otimes \Delta g \right]
\end{eqnarray}
where 
\[ \Delta q_{NS}=\sum_{i=1}^{n_f}\left(\frac{e_i^2}{<e^2>}-1\right)
   (\Delta q_i +\Delta \overline{q_i})~{\rm and} \]
\[ \Delta q_S=\sum_{i=1}^{n_f}(\Delta q_i+\Delta \overline{q_i}).\]
and they evolve according to the Altarelli-Parisi equations:
\begin{eqnarray}
\label{eq:glap}
\frac{\id }{\id \ln t}\Delta q_{NS}&=&\frac{\alpha_s(t)}{2\pi}\Delta
                              P^{qq}_{\rm NS}\otimes\Delta q_{NS}\\
\nonumber\\
\frac{\id}{\id \ln t}
       \left(
          \begin{array}{c}
             \Delta q_S\\
             \Delta g\\
            \end{array}
          \right)
       &=&\frac{\alpha_s(t)}{2\pi}
       \left(
          \begin{array}{cc}
             \Delta P_{\rm S}^{qq} & 2n_f \Delta P_{\rm S}^{qg}\\
             \Delta P_{\rm S}^{gq} &  \Delta P_{\rm S}^{gg}
             \end{array}
          \right) \nonumber \\
     & & \otimes
       \left(
           \begin{array}{c}
              \Delta q_S\\
              \Delta g
              \end{array}
           \right),
\end{eqnarray}
where $t=\ln (Q^2/\Lambda^2)$. The splitting functions $\Delta P^{ij} \equiv
P^{i^+j^+}-P^{i^+j^-}$ at next-to-leading order,
\[ P(x,Q^2)=P^{(0)}(x)+\frac{\alpha_s(Q^2)}{2\pi} P^{(1)}(x), \]
have recently been calculated \cite{mertig}. Differences between spin-dependent
and spin-averaged splitting functions give rise to different $Q^2$-evolutions
of $g_1$ and $F_1$. Therefore the assumption made in most of the analyses
up to now that $A_1=g_1/F_1$ is independent of $Q^2$, is strictly not valid.

\subsection{Extrapolation of $g_1$ in $x$}
Extrapolations of $g_1(x,Q_0^2)$ at large $x$, 
from $x_{max} \sim 0.7$ to $x=1$, were made
assuming $g_1(x) \sim F_1(x)$ or $\sim (1-x)^3$ or asumming 
$A_1(x\rightarrow 1)\rightarrow 1$. The resulting contributions to $\Gamma_1$
are very small and therefore this subject is not controversial.

However, extrapolation at small $x$, from $x_{min}$=0.003 
for the SMC data and
$\sim$0.03 for experiments at SLAC and DESY to $x=0$ requires fenomenological
input for the behaviour of the dominant singlet contribution. This behaviour 
depends on the interplay between expectations from
\begin{itemize}
\item Regge trajectories $(a_1$ and $f_1)$, 
$g_1(x) \sim x^{\alpha},~0\leq \alpha \leq 0.5$, used by the SMC, E142, E143
and HERMES,
\item non-perturbative Pomeron \cite{landshoff}, $g_1(x) \sim \log(1/x)$,
alternatively used by the E143,
\end{itemize}
and the expectations from perturbative QCD which were not used in
the experimental analyses:
\begin{itemize}
\item $|g_1(x)|$ rises faster than $\sim \log(1/x)^{\nu}$ for any 
$\nu$ and slower than $\sim 1/x^{\lambda}$ for any $\lambda > 0$ \cite{bfr}
\item $g_1(x) \sim 1/x^{\omega}$ with $\omega > 1$ for $\alpha_s > 0.12$,
where a strong rise of $g_1(x)$ at small $x$ is possible \cite{bartels}.
\end{itemize}
Experiments assign a 100\% error to their extrapolations at small $x$ but
the question of underlying assumptions is not settled; measurements
of $g_1$ at smaller $x$ are needed. The need to have small extrapolation error is
clearly visible in very high statistics E154 data on $g_1^n$.

\subsection{Higher-twist terms}
In the operator product expansion $\Gamma_1$ is given by
\begin{eqnarray}
\label{eq:HT}
\Gamma_1(Q^2) & = & \Gamma_1^{LT}(Q^2)+ \nonumber \\
   & & \underbrace{\frac{m^2}{9Q^2}\left[a^{(2)}+4d^{(2)}+4f^{(2)}\right]}
        _{C_{HT}/Q^2}+ \nonumber \\
   & & {\cal O}\left(\frac{m^4}{Q^4}\right)
\end{eqnarray}
where $a^{(2)}$, $d^{(2)}$ and $f^{(2)}$ are the matrix elements of  
twist-2, twist-3 and twist-4 operators, respectively.

There are first experimental results\cite{g2pd} from E143 
on $a^{(2)}_{p(d)}$ and $d^{(2)}_{p(d)}$. They require
$g_2(x)$ as input and have large statistical errors.

Phenomenological estimates of their magnitude come from QCD sum rules,
renomalon methods, lattice QCD and
from MIT bag model (see Ref. \cite{mank} for a review and also Ref. \cite{ht_new}). 
Their values differ, even in sign. However,
the implied corrections to $\Gamma_1$ are in general smaller than 
the experimental errors. They are also less important at higher $Q^2$.

\subsection{Results for $\Gamma_1$}
Recent experimental results for $\Gamma_1$ are shown in Table 3.
They were obtained under the assumption that $A_1$ is independent of $Q^2$
and using Regge-type extrapolations of $g_1$ to $x=0$.
\begin{table*}[t]\begin{center}
\caption{Experimental results for the first moments of the spin-dependent
structure functions. The values in parantheses give the statistical and
the systematic error, respectively. Except for the E143, all results are 
preliminary. E142 results are from Ref.$^{17}$.}
\vspace{0.4cm}
\begin{tabular}{|l|c|l|l|l|}
\hline 
\multicolumn{1}{|l}{Experiment} 
&\multicolumn{1}{|c}{$Q^2$} 
& \multicolumn{1}{|c}{$\Gamma_1^{\rm p}$} 
& \multicolumn{1}{|c}{$\Gamma_1^{\rm d}$} 
& \multicolumn{1}{|c|}{$\Gamma_1^{\rm n}$}\\ 
&{$({\rm GeV}^2)$} & & & \\ \hline 
SMC    & 10 & 0.136(14)(\space 9) & 0.038(7)(5) 
& $-$0.055\space \space (24)$^{a)}$\\
E143   &  3 & 0.127(\space 4)(10) & 0.042(3)(4) 
& $-$0.037(\space 8)(11)$^{a)}$\\
E142   &  2 & ---          & ---                
& $-$0.031(\space 6)(\space 9)\\
HERMES &  3 & ---          & ---                
& $-$0.032(13)(17)\\
E154   &  5 & ---          & ---                
& $-$0.037(\space 4)(10)$^{b)}$\\
\hline 
\mco{5}{l}{$^{(a)}$ from deuteron and proton data.} \\
\mco{5}{l}{$^{(b)}$ in the measured range only ($0.014 < x < 0.7$).} \\
\end{tabular}
\end{center}
\end{table*}

\section{Physics interest in $\Gamma_1$}
The first moments of the spin-dependent structure functions $g_1^p$
and $g_1^n$ are used to test 
the fundamental Bjorken sum rule \cite{bjsr}. 
Furthermore, $\Gamma_1$ is used to evaluate the singlet element of the axial-vector current, $a_0$. In the leading-order it is equal to the total
spin carried by quarks, $a_0=\Delta\Sigma$, and its value was predicted
using the Ellis-Jaffe sum rule \cite{ejsr}. The value found
by the EMC \cite{emc} was much smaller than the prediction. This discovery has led to 
the "spin crisis" and triggered
large experimental and theoretical activity.

\subsection{Bjorken sum rule}
Under the assumption of the isospin symmetry between neutron and proton
we can write for the axial charges:
\[ a_u^n=a_d^p,~a_d^n=a_u^p,~{\rm and}~a_q^n=a_q^p;~q=s,c,b,t. \]
The singlet term cancels in the difference $\Gamma_1^p-\Gamma_1^n$ 
(see Eq.9) and we obtain the Bjorken sum rule (BSR) prediction:
\begin{equation}
(\Gamma_1^p-\Gamma_1^n)(Q^2)=\frac{1}{6}C_{NS}(\alpha_s(Q^2)) \cdot
|g_A|.
\end{equation}
This is a very accurate prediction: the value of $|g_A|$, the coupling
constant in neutron beta-decay, is known to within 2 permille \cite{pdb} 
and the perturbative expansion of $C_{NS}$ is known to ${\cal O}(\alpha_s^3)$
\cite{larin_ns} and there is also an estimate of ${\cal O}(\alpha_s^4)$ term
\cite{o4term}. 
Therefore it is regarded as a test of perturbative QCD.
Conversely, since the perturbative expansion is sensitive to the value
of $\alpha_s$, the BSR can be used to determine $\alpha_s(M_Z^2)$.

Theoretical uncertainties were discussed above; they are due to the assumptions
used to obtain $\Gamma_1^{p(n)}$ and to the contributions from
higher-twists, $C_{HT}^{p-n}/Q^2$.

\subsection{Experimental results for the Bjorken sum}
The experimental results for the Bjorken sum are given in Table 4.
They are compared to the predictions in Figure 12 where we used
$n_f=3$ and $\alpha(M_Z^2)=0.117 \pm 0.005$. 
Good agreement with the prediction is observed.
\begin{table}\begin{center}
\caption{Experimental results for the Bjorken sum. Except for the E143, all results
are preliminary.
The value in parathesis gives the total error. For $\Gamma_1^n$ from E142 we use result from 
Ref.$^{17}$}
\vspace{0.4cm}
\begin{tabular}{|c|c|c|} \hline
& &  \\
Experiment & $Q^2$ & $\Gamma_1^p-\Gamma_1^n$ \\ 
           &{$({\rm GeV}^2)$} & \\ \hline 
SMC      & 10 & 0.191(36) \\
E143     &  3 & 0.163(19) \\
E143/E142&  3 & 0.159(14) \\ \hline
\end{tabular}
\end{center}
\end{table}
\begin{figure}%[h]
\begin{tabular}{p{0.45\textwidth}}
\vskip-3.0cm
\epsfig{file=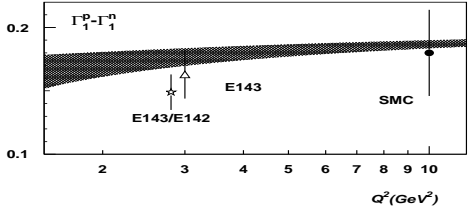,height=10.0cm,width=\linewidth}
%,clip=
%,bbllx=5pt,bblly=5pt,bburx=530pt,bbury=530pt} 
%\vskip-0.3cm
\caption{Results for Bjorken sum from recent experiments. The shaded area
indicates the predictions.}
\end{tabular}
\vskip-0.5cm
\end{figure}
%\vskip-0.5cm

\subsection{Determination of $\alpha_s$}
Figure 12 shows also the sensitivity of the predictions to 
the value of $\alpha_s$; the spread in the predictions is mainly due to 
its error.  It is therefore tempting to derive the value
of $\alpha_s(M_Z^2)$ from the experimental results on the Bjorken sum,
assuming that the prediction given by Eq.16 holds precisely. Since
all known terms of the perturbative expansion are negative such 
procedure requires an estimate of the contribution from all higher
order terms. The most recent determination \cite{egks} of $\alpha_s$
from BSR relies on Pad\'{e} summation: using an estimate 
of the average experimental value 
$(\Gamma_1^p-\Gamma_1^n)(3\,{\rm GeV}^2)=0.160 \pm 0.014$ and
$C_{HT}^{p-n}=-0.02 \pm 0.01$, the authors obtain
\[ \alpha_s(M_Z^2)=0.117^{+0.004}_{-0.007}~(exp.)~\pm~0.002~(th.). \]
The theoretical error is mainly due to a renormalisation scale dependence 
and it is very small.

\subsection{Results for $a_0$ and for the Ellis-Jaffe sum  rule}
The value of the singlet axial charge ($a_0$) is obtained from Eq.(9) using
experimental values for $\Gamma_1$; $a_{NS}$ is calculated from 
$g_A$ and from $F$ and $D$ 
values \cite{fd} using Eq.(10) and $C_S$ has been evaluated \cite{larincs} 
up to
${\cal O}(\alpha_s^2)$. It
can be written as \cite{larincs}:
\[ a_0(Q^2)=[1+ c_1 \alpha_s(Q^2) + c_2 \alpha_s^2(Q^2)] \cdot
(a_0)_{inv.}\]
where $c_i$ are numerical constants and $(a_0)_{inv} \equiv a_0(Q^2 \rightarrow \infty)$.
Using $a_0$ interpretation from the $\overline{MS}$-scheme (Eq.(11)),
$a_0(Q^2) \equiv \Delta\Sigma(Q^2)$ is the total spin carried by all quarks.
With this interpretation and under the assumption of the flavour
symmetry in the polarised sea, the value of $a_s \equiv \Delta s$ can be calculated.

In the leading-order $\Delta\Sigma$ is invariant. Its value
can be predicted assuming Ellis-Jaffe assumption that OZI rule is valid ($\Delta s=0$):
$\Delta\Sigma=\Delta u +\Delta d=a_8=3F-D \simeq 0.6$.\\
The values of $(\Delta\Sigma)_{inv.}$ and $\Delta s$ are given in Table 5.
\begin{table}[t]
\begin{center}
%\vskip-2.8cm
\caption{Results for $(\Delta\Sigma)_{inv.}$ and for $\Delta s$.}
\vspace{1.0cm}
\begin{tabular}{|lc|l|l|}
\hline 
\multicolumn{2}{|l}{Experiment} 
& \multicolumn{1}{|c}{$(\Delta\Sigma)_{inv.}$} 
& \multicolumn{1}{|c|}{$\Delta s $} \\
&  & & \\ \hline 
SMC         & p & 0.21 (16)      & $-0.12$ (5) \\ 
            & d & 0.24 (\space 8)& $-0.11$ (3) \\ \hline
E143        & p & 0.24 (10)      & $-0.11$ (3) \\
            & d & 0.30 (\space 5)& $-0.09$ (2) \\ \hline
%E142        & n & 0.47 (10)      & $-0.03$ (3) \\ 
%\hline
\end{tabular}
\end{center}
\end{table}
\begin{figure}%%[h]
\begin{tabular}{p{0.45\textwidth}}
%\vskip3.0cm
\epsfig{file=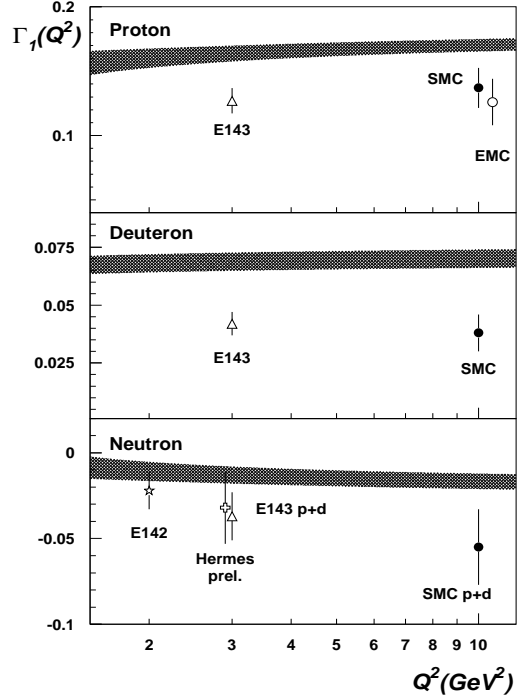,height=9cm,width=\linewidth
%,clip=
,bbllx=25pt,bblly=20pt,bburx=420pt,bbury=550pt} 
%\vskip-0.3cm
\caption{Comparision of the first moments $\Gamma_1$ to the Ellis-Jaffe
sum rule prediction. Also plotted is the first result from the EMC which gave 
rise to "the spin crisis".}
\end{tabular}
\vskip-1cm 
\end{figure}
%\end{table}
%\vskip3.5cm
Clearly, $\Delta \Sigma$ is below expectation and the OZI rule is violated.
%\vskip3cm
An alternative way to demonstrate this discrepancy is to compare $\Gamma_1$ 
to the value predicted under the assumption of  $\Delta s=0$ 
(Ellis-Jaffe sum rule). With this assumption we have $a_0=a_8$ and
\begin{eqnarray}
(\Gamma_1^{p(n)})_{EJ}(Q^2)&=
   & C_{NS}(Q^2)\left(\pm\frac{1}{12}a_3+\frac{1}{36}a_8\right)+ \nonumber \\
   & &\frac{1}{9}C_S(Q^2)a_8.
\end{eqnarray}
Figure 13 clearly shows that $a_0$ should be bigger than $a_8$ which requires
$a_s<0$. 
%\vskip2cm
\subsection{Present understanding of $a_0$}
We can rewrite Eq.(11) in the form:
\[ a_0(Q^2) = \left\{ \begin{array}{ll}
 \underline{\overline{\rm{MS}};~\Delta G~ {\rm enters~via~polarised~sea}:} \\
          \Delta\Sigma(Q^2) \\
          \\
 \underline{{\rm AB};~\Delta G~{\rm enters~explicitely}:}\\
          \Delta\Sigma  -\underbrace{
     n_f \frac{\alpha _s(Q^2)}{2\pi} \Delta G(Q^2)}_{\sim Q^2-{\rm independent}}

                        \end{array}
                \right. \]
\begin{equation} \end{equation}
In the AB-scheme, where gluon polarisation enters explicitely, one can 
estimate $\Delta G$ required in order to obtain $\Delta \Sigma=0.6$;  from
Eq.18 we obtain $\Delta G(10\,{\rm GeV}^2) \sim 2$.

To summarise: the physics interpretation of $a_0$ in the NLO is 
scheme-dependent. Its value is below the expectation from the LO due
to gluon polarisation.
%\vskip2cm
\section{Determinations of $\Delta G$}
Indirect determinations of $\Delta G$ come from fits of parametrised
polarised parton distributions to all $g_1^{p(d,n)}$ data, 
using the Altarelli-Parisi
evolution equations (Eqs.13 and 14) in the NLO. There are three
independent analyses \cite{bfr} \cite{gerstir} \cite{grsv} and in the range 
$1 < Q^2 < 10\,{\rm GeV}^2$ they all require $\Delta G  \sim 1 \div 2$.

Sizeable gluon polarisation clearly affects the $Q^2$-dependence of $g_1$.
Since the calculation of  $\Gamma_1(Q_0^2)$ requires 
the extrapolations of $g_1(x,Q^2_{meas.}) \rightarrow g_1(x,Q^2_0)$,
it might lead to different value of $\Gamma_1$. Figures 14 and 15
show the preliminary results from the NLO QCD fits made by the SMC
to the recent E143 and SMC data on $g_1^p$ and $g_1^d$
using the procedure of Ref. \cite{bfr}. Lines show the pattern of
the $Q^2$-dependence ; the solid line shows fitted
$g_1$ at $Q^2$ measured by the SMC, dashed and dotted lines are 
the values at $Q^2$ of $1\,{\rm GeV}^2$ and $10\,{\rm GeV}^2$, respectively.
\begin{figure}%[ht]
\vskip-0.5cm
%\begin{tabular}{p{0.45\textwidth}}
\epsfig{file=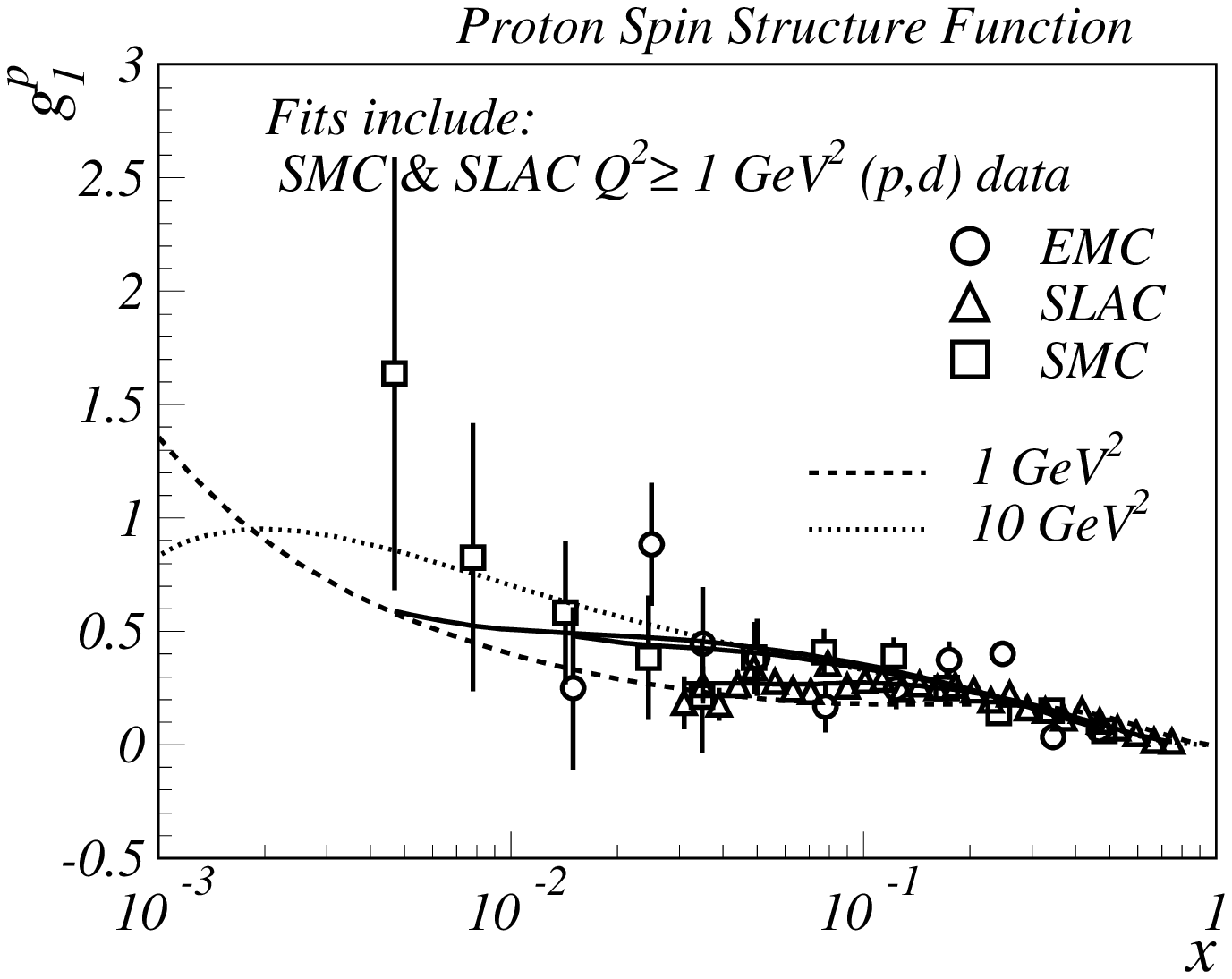,height=4cm,bbllx=0pt,bblly=50pt,bbury=331pt,bburx=450pt}
%height=10.0cm,
%,clip=
%,bbllx=5pt,bblly=5pt,bburx=530pt,bbury=530pt} 
%\vskip-3.3cm
\caption{$g_1^p$ from the NLO QCD fits to $g_1^p(x,Q^2)$ and $g_1^d(x,Q^2)$
from the E143 and the SMC. Points show $g_1^p(x)$ at measured $Q^2$. See
text for explanations.}
%%\end{tabular}
%%\end{figure}
%\vskip-0.5cm
%%\begin{figure}%%[h]
%%\begin{tabular}{p{0.45\textwidth}}
%\vskip-1.0cm
\epsfig{file=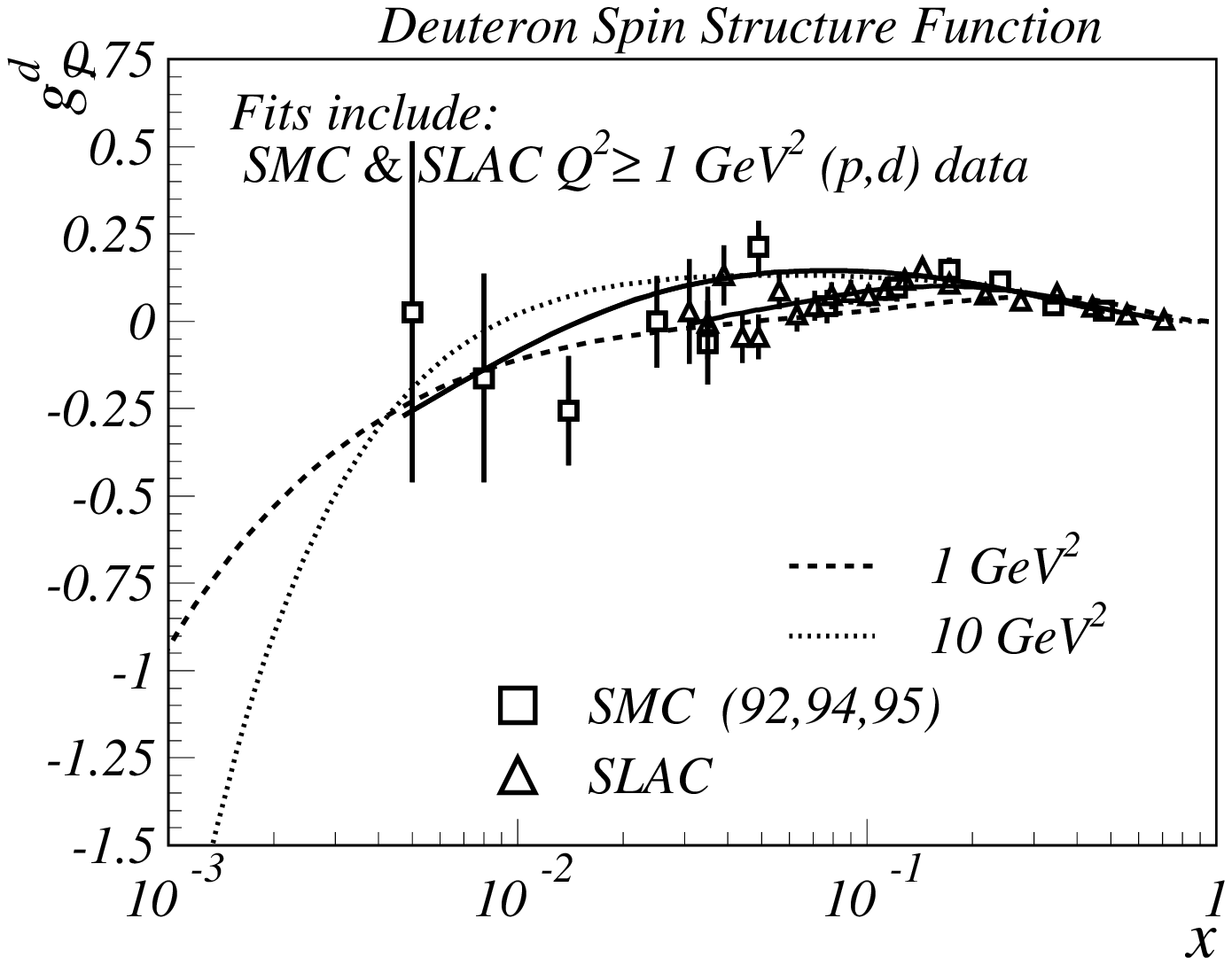,height=4cm,bbllx=0pt,bblly=50pt,bbury=331pt,bburx=450pt}
%,clip=
%,bbllx=5pt,bblly=5pt,bburx=530pt,bbury=530pt} 
%\vskip-0.3cm
\caption{$g_1^d$ from the NLO QCD fits to $g_1^p(x,Q^2)$ and $g_1^d(x,Q^2)$
from the E143 and the SMC. Points show $g_1^d(x)$ at measured $Q^2$. See
text for explanations.}
%\end{tabular}
\vskip-0.5cm
\end{figure}
%\vskip-0.5cm
Using the extrapolation
%\vskip3cm
\begin{eqnarray}
g_1(x,Q_0^2)&=&g_1(x,Q^2_{meas.}) + \nonumber \\
  &&[g_1(x,Q_0^2)-g_1(x,Q^2_{meas.})]_{\rm from~fit} \nonumber
\end{eqnarray} \begin{equation} \end{equation}
the SMC obtains (preliminary) at $Q_0^2=10\,{\rm GeV}^2$:
\begin{itemize}
\item $\Gamma_1^p=0.130~~~(0.136\pm 0.017)$
\item $\Gamma_1^d=0.037~~~(0.038\pm 0.009)$
\end{itemize}
where the values given in parantheses are from Table 3 and
they were evaluated assuming that $g_1/F_1$ does not
depend on $Q^2$.\\
The theoretical error of $\Gamma_1$ due to uncertainty in 
$\Delta g(x,Q^2_0)$ and to unknown higher-order corrections was estimated 
\cite{bfr} to be $^{+0.009}_{-0.005}$.

\section{Results on the polarised quark distributions}
First results on the polarised quark distributions were obtained by the SMC
\cite{had} from the analysis of asymmetries of semi-inclusive and 
inclusive cross sections. There are new (preliminary) results from 
the SMC which include new deuteron data.

The results are obtained assuming QPM.
The asymmetry of inclusive  virtual photon absorption 
cross sections 
is written in terms of the distribution functions of quarks ($q(x,Q^2)$)
and of their spin ($\Delta q(x,Q^2)$):
\begin{equation}
A_1(x,Q^2)=\frac{\sum_q e_q^2\,\Delta q(x,Q^2)}{\sum_q e_q^2\,q(x,Q^2)}.
\end{equation}
The asymmetry of semi-inclusive cross sections depends also on 
the fragmentation functions of  quarks $q$ into a final state hadron $h$,
$D_q^h(z,Q^2)$, where $z$ is the fraction of quark's energy 
taken by a hadron. The SMC has determined semi-inclusive asymmetries for
the positive and negative hadron yields integrated over $0.2<z<1$, 
$D_q^h(Q^2)=\int_{0.2}^{1}D_q^h(z,Q^2){\rm d}z$:
\begin{equation}
A_1^{+(-)}(x,Q^2)=\frac{\sum_{q,h} e_q^2\,\Delta q(x,Q^2)\,D_q^h(Q^2)}
                         {\sum_{q,h} e_q^2\,q(x,Q^2)\,D_q^h(Q^2)}.
\end{equation}
In Eqs.20 and 21 quark distributions were taken from the MRS parametrisation
\cite{mrs} and the fragmentation functions of non-strange quarks into pions 
were obtained from
the EMC measurements \cite{ffunct} by using charge conjugation and isospin
 symmetry.
In each bin of $x$ these equations for proton and deuteron targets 
constitute a system of $2+4$
linear equations written in terms of three unknown spin distribution:
of valence quarks, $\Delta u_v(x)$ and $\Delta d_v(x)$, and of the sea 
antiquarks, $\Delta \overline q(x)$.
They were evaluated by the least-square method and the results
are shown in Fig.16.
\begin{figure}%%[h]
\begin{tabular}{p{0.45\textwidth}}
%\vskip-0.5cm
\epsfig{file=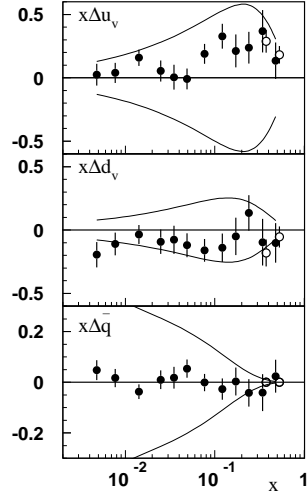,height=10.0cm,width=\linewidth}
%,clip=
%,bbllx=5pt,bblly=5pt,bburx=530pt,bbury=530pt} 
%\vskip-0.3cm
\caption{Quark spin distribution functions 
(a) $x \Delta {u}_v(x)$, (b) $x \Delta {d}_v(x)$, (c) $x \Delta \bar{q}(x)$.
The open circles are obtained when the sea polarisation is set to zero while the closed circles are obtained without
this assumption.
The error bars are statistical. The curves correspond to the upper and the lower limits $\pm q(x)$
from the unpolarised quark distributions  evaluated at $Q^2=10$~GeV$^2$.}
\end{tabular}
\end{figure}
We observe that $\Delta u_v(x)$ is positive while $\Delta d_v(x)$ is negative.
The spin distribution of the non-strange sea $\Delta \bar q(x)$ is compatible with zero even at small $x$ where the unpolarised sea is large. 
There is an indication of
a non-vanishing valence quark polarisation $\Delta q_v/q_v$ at small $x$ which
is consistent with the observation of $g_1^n \neq g_1^p$
at the smallest $x$ in the SMC data.

\section{How the proton spins?}
In terms of the nucleon constituents we expect:
\begin{equation}
\frac{1}{2} = \frac{1}{2}\Delta \Sigma + \Delta G(Q^2) + L_z(Q^2),
\end{equation}
where the $Q^2$-evolution of $\Delta G(Q^2)$ and $L_z(Q^2)$, the orbital 
angular momentum of quarks and gluons, must be coupled. This issue 
raises theoretetical activity \cite{ji} but it is still under discussion. 
As discussed in previus sections, the OZI rule gives $\Delta \Sigma \sim 0.6$,
which requires $\Delta G(Q^2=10\,{\rm GeV}^2) \sim 2$ and the equation 
above leads to $\L_z \sim -1.8$. 

\section{Prospects for polarised deep-inelastic scattering}
\subsection{New results from the approved experiments}
In the near future we can expect new results from the following experiments:
\begin{itemize}
\item SMC will have more data on $g_1^p$ and $\Delta q$. There might be new
results for $Q^2<1\,{\rm GeV}^2$ and $x<0.003$,
\item E155 will have new high precision data on $g_1^{p(d)}$ for 
$Q^2>1\,{\rm GeV}^2$ and $x>0.014$,
\item HERMES will have new data on $\Delta q$ (with identified hadrons)
and also on $g_1^{p(n,d)}$.
\end{itemize}
These data will certainly increase the precison of the present results
and might produce new, interesting observations. However they will
not solve two outstanding problems which came up recently,
\begin{enumerate}
\item to determine the shape of $g_1(x)$ at smaller $x$ and $Q^2>1\,{\rm GeV}^2$,
\item to determine directly gluon polarisation $\Delta g(x)$,
\end{enumerate}
which can be solved only by new experiments.

\subsection{Proposed experiments}
\subsubsection{COMPASS at CERN}
Experiment has been proposed on the muon beam line at CERN and recently 
recommended by the SPSL Committee. The main subject of the deep-inelastic
scattering part of the  program
is to determine $\Delta g/g$ from asymmetries of cross sections for
open charm production, $D^0(\overline D^0)$:
\begin{equation}
A_{\gamma N}^{c\bar{c}}(\nu)=
\frac{\int_{4m_c^2}^{2M\nu}{\rm d}\hat{s}\,\Delta
\sigma(\hat{s})\,\Delta \G(x_g,\hat{s})}
{\int_{4m_c^2}^{2M\nu}{\rm d}\hat{s}\,\sigma(\hat{s})\,\G(x_g,\hat{s})}.
\label{eq:acc}
\end{equation}
where $\hat{s}=(q+k)^2$, $x_g=\hat{s}/2M\nu$ denotes the nucleon momentum 
fraction carried by the gluon and $\Delta \sigma$ is the spin-dependent part of
the charm production cross section in the photon-gluon process, shown in
Figure 17.
\begin{figure}%%[h]
\begin{tabular}{p{0.45\textwidth}}
%\vskip-0.5cm
\epsfig{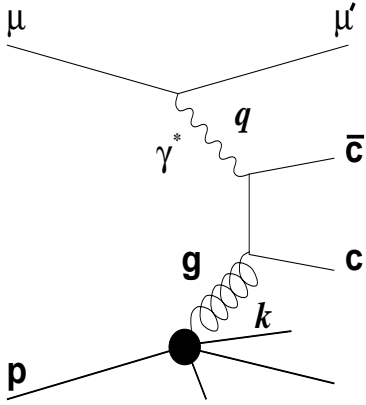}
%,clip=
%,bbllx=5pt,bblly=5pt,bburx=530pt,bbury=530pt} 
%\vskip-0.3cm
\caption{The photon-gluon fusion diagram.}
\end{tabular}
\end{figure}
Using 100~GeV $\mu$-beam the experiment will cover $x_g>0.07$ where the accuracy
$\delta A^{c\bar c} \sim 0.05$ is expected. The corresponding accuracy
in gluon polarisation is $\delta(\Delta G/G) \sim 0.14$. The experiment
is expected to start data taking around the year 2000.

Similar experiment on the electron beam at SLAC is being discussed.

\subsubsection{At polarised HERA}
At HERA the polarised deep-inelastic physics is feasible \cite{kunne}
with the present H1 and ZEUS detectors if the integrated
luminosity $\int_{}^{} {\cal L} {\rm d}t \sim 200\,pb^{-1}$ is reached for electron
and proton polarisations $\sim 0.7$. Three physics items
are considered.\\

{\underline {1. $\Delta g$ from di-jet (2+1) events:}}\\
These events are produced via photon-gluon fusion process shown in Figure 17,
where two jets arise from quark pairs. The momentum fraction carried
by the gluon can be determined from $\hat s$, $x_g=x(1+\hat s/Q^2)$, where
$x$ is the Bjorken`s variable. The expected asymmetries of cross sections are 
$\sim 0.03$ and they can be measured to a high accuracy \cite{feltesse}.\\

{\underline {2. $\Delta g$ from $Q^2$-evolution of $g_1(x,Q^2)$}:}\\
The measurements cover $x \geq 10^{-4}$ with $Q^2$ bigger by a factor
of 100 compared to fixed-target experiments.  However the expected
asymmetries are small ($\sim 5\cdot 10^{-4}$) and the kinematic depolarisation 
factor is unfavourable.\\

{\underline {3. Polarisation of quarks from $g_5^{{\rm W}^-({\rm W}^+)}:$}}\\
It has been proposed \cite{kalinowski} to study polarised 
charge-current processes using both $e^-p$ and $e^+p$ interactions 
in order to determine new polarised structure functions 
$g_5^{{\rm W}^-({\rm W}^+)}$. They are expressed by different flavour
combinations compared to $g_1$, e.g. 
$g_5^{\rm W^-}+g_5^{\rm W^+} =\Delta u_{\rm v} + \Delta d_{\rm v}$.
For $x>0.1$ very large asymmetries (bigger than 0.1) are expected.

The feasibility to have polarised protons at HERA is under study.
The possible time scale is most likely to be beyond year 2004.

\section{Conclusions}
There is a good agreement between experiments at SLAC, CERN and DESY
on the results on $g_1^{p(d,n)}$. \\
The Bjorken sum holds well and is being
used to determine $\alpha_s$ which might have small theoretical error.

New results are consistent with the singlet axial charge $a_0 \simeq 0.25$ 
which is below the expectation from the OZI rule and corresponds
to $a_s \simeq -0.1$.

On the theoretical side, there is a large progress in understanding this 
effect in terms of the polarisation of gluons, 
$\Delta G \sim 1 \div 2$ at $Q^2 \sim 10~{\rm GeV}^2$. However
with the present data is not sufficient to determine it more 
precisely from the scaling violations alone. It might be possible with
the new data coming from the SMC, E155 and HERMES.

The COMPASS experiment has been recommended at CERN to determine the gluon 
polarisation from the open charm production in the photon-gluon fusion process.
The option to have polarised protons at HERA is under the study and
several processes have been proposed to investigate the gluon polarisation 
there.

\section*{Acknowledgements}
We thank the Organising Committee of the XXVII International Conference 
on High Energy Physics for the invitation to present this talk.
This work was supported by the Polish Committee for Scientific Research
(KBN) SPUB/P03/112/95 and SPUB/P03/114/96.

\footnotetext[1]{Plenary talk presented
at the XXVII International Conference on High Energy Physics,
25-31 July 1996, Warsaw, Poland.}

\end{document}